# CRUDE OIL AND MOTOR FUEL: FAIR PRICE REVISITED

Ivan O. Kitov[1], Oleg I. Kitov

[1]Institute for the Geospheres' Dynamics, Russian Academy of Sciences

**Abstract**
In April 2009, we introduced a model representing the evolution of motor fuel price (a subcategory of the consumer price index of transportation) relative to the overall CPI as a linear function of time. Under our framework, all price deviations from the linear trend are transient and the price must promptly return to the trend. Specifically, the model predicted that "*the price for motor fuel in the US will also grow by 50% by the end of 2009. Oil price is expected to rise by ~50% as well, from its current value of ~$50 per barrel.*" The behavior of actual price has shown that this prediction is accurate in both amplitude and trajectory shape. Hence, one can conclude that the concept of price decomposition into a short-term (oscillating) and long-term (linear trend) components is valid. According to the model, the price of motor fuel and crude oil will be falling to the level of $30 per barrel during the next 5 to 8 years.

## Introduction

In the beginning of 2009 we developed a model [1, 2] predicting the long-term price evolution for various subcategories of consumer and producer price indices as well as major commodities: gold, crude oil, metals, etc. The model was based on one prominent feature of the difference between consumer (producer) prices of individual components and the overall consumer (producer) price index. These differences are characterized by the presence of sustainable long-term (quasi-) linear trends. For many producer price indices, these trends are slightly nonlinear but still robust. They are observed in subcategories with varying weights in the CPI and PPI: meats [3], gold ores [4], durables and nondurables [5], jewelry and jewelry related products [6], and motor fuel [7].

For major subcategories these trends last between five and twenty years and then turn to trends with opposite slopes. The transition to new trends lasts three years at most. However, there are subcategories without slope changes as reported by the Bureau of Labor Statistics [8], where all CPI and PPI time series were retrieved from. The best example of such a one-leg trend since 1980 is the price index of medical care [1]. The index of communication has been linearly deviating from the headline CPI since 1998 (in this study we use seasonally adjusted CPIs and not seasonally adjusted PPIs), i.e. since the beginning of reporting; before it had been reported as an indistinguishable part of the index of education and communication.

In the short run, actual prices oscillate around the long-term trends with varying amplitudes. In a sense, the trends represent the lines of gravity centers for given prices and any large deviation from the trends must be compensated promptly. As a result, both short- and long-term predictions of commodity prices are feasible. In the long run, the prices follow up the trends. In the short-run, the next move in a given price depends on the current position relative to corresponding trend. When very far from the trend, the price is more likely to start returning. When approaching the trend, the price may choose any direction for the further evolution, i.e. it should not inevitably go the other side of the trend. Using long-term trends and short-term deviations we predicted the evolution of prices for gold, durables and nondurables, jewelry and a number of consumer price indices. These predictions will be revisited in due course. In this paper, we focus on crude oil and motor fuel.

For the price index of motor fuel, Kitov and Kitov [7] developed a similar model as based on the deviation from the core CPI, i.e. the headline CPI less food and energy. Using this model, we predicted the evolution of oil price as well. The overall performance of the model between March and December 2009 was reported in [9]. Here we also revise the long-term prediction of crude petroleum and motor fuel price and make necessary corrections to the model as related to the observations since March 2009.

## 1. The model

The model derived in [1, 2] implies that the difference between the overall CPI (same for the PPI), *CPI (PPI)*, and a given individual price index iCPI (iPPI), can be described by a linear time function over time intervals of several years:

$$CPI(t) - iCPI(t) = A + Bt \quad (1)$$

, where $A$ and $B$ are the regression coefficients, and $t$ is the elapsed time. Therefore, the "distance" between the CPI and the studied index is a linear function of time, with a positive or negative slope $B$. Free term $A$ compensates the difference related to the start levels for a given year. For example, the index of communication was started from the level of 100 in December 1997 when the overall CPI was already at the level of 161.8 (base period 1982-84 =100).

Figure 1 displays examples of linear trends in the two differences related to the scope of this paper. In the left panel, the evolution of the index of motor fuel relative to the headline CPI is shown. Notice that in the original paper [7] we referred the index of motor fuel to the core CPI, but the discrepancy between the headline and core CPI is negligible relative to the change in the index of motor fuel. There are two distinct periods of linear dependence on time: from 1980 to 1999 and from 2001 to 2008. Apparently, there is one finished transition period between 1999 and 2001, where the trend with a positive slope ($B$=+4.2) changed to a negative one ($B$=-21.1), in both cases the goodness-of-fit being very high: $R^2$~0.9. The first transition period is characterized by elevated price volatility. Since 2008, the negative trend in the difference has been suffering a transition to a positive one, which is shown in Figure 1 by a dashed line. This transition is characterized by a much higher volatility and has been fading away since the end of 2009. A new trend has been emerging since the beginning of 2010. Without prejudice, we drew the new trend as a mirror reflection to the previous one. Supposedly, it will last seven years. As a result, the difference will increase from -50 in 2009 to +75 in 2016 since the index of motor fuel will be falling at a rate of 17.9 units per year relative to the headline CPI. This casts the long-term prediction of the motor fuel index.

In the right panel, the difference between the PPI and the index of crude petroleum (domestic production) is shown between 1985 and 2010. Expectedly, there are two distinct periods of linear dependence on time: from 1988 to 1999 and from 2001 to 2008. The regression coefficients in both periods are different from those for the index of motor fuel: +2.9 and -17.1, respectively. In other words, the motor fuel index magnifies the change in oil price. There was one transition period between 1999 and 2001, where the original positive trend was turned down. Even a simple visual inspection of the transition period reveals several differences in timing and amplitude between motor fuel and crude oil: the former started to fall three months later and came to the level of 1999 in the end of 2001.  The index of crude oil fell by ~15 units relative to its level in 1999.

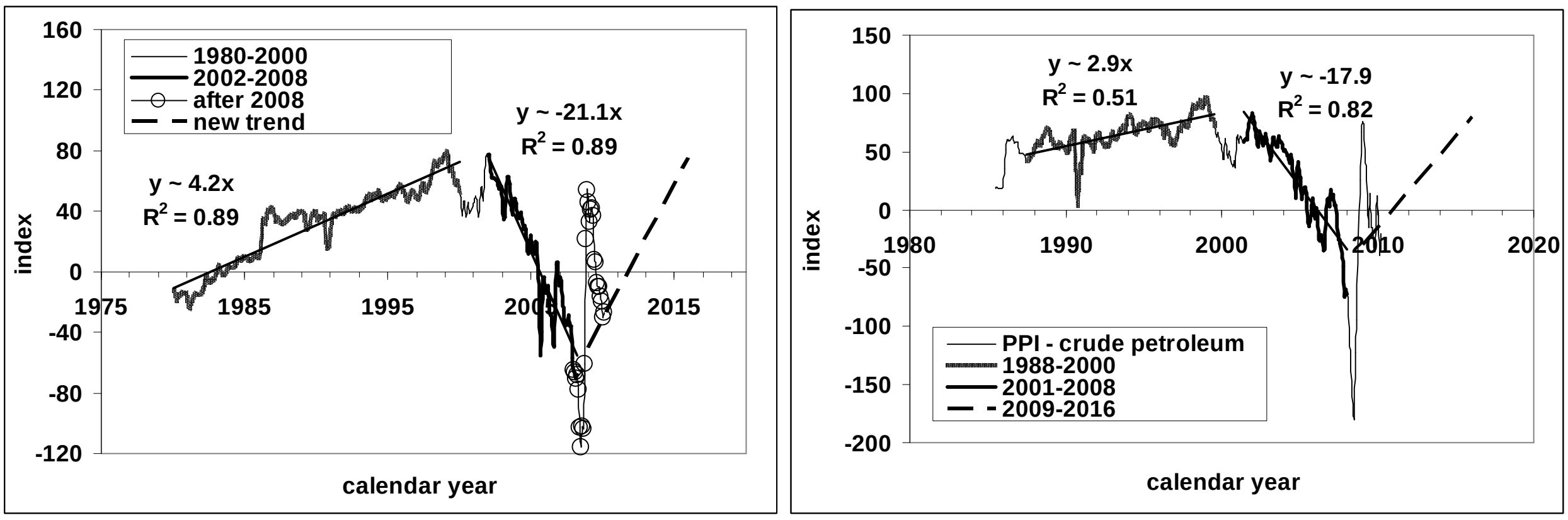

Figure 1. Illustration of linear trends. *Left panel*: the difference between the headline CPI and the index of motor fuel between 1980 and 2010. *Right panel*: The difference between the overall PPI and the (producer price) index of crude petroleum (domestic production). In both panels: there are two quasi-linear segments with a turning point near 2000. Since the end of 2008, both differences have been passing a transition. Linear trends with relevant linear regression lines and corresponding slopes are also shown.

From Figure 1 (and many others published before), one can conclude that the presence of linear trends is a basic feature of the CPI and PPI. Another fundamental characteristic of the differences consists in the fact that all deviations from the trends were only short-term ones. This implies that any current or future deviations from the new trends in Figure 1, which have been under development since 2008, must be compensated promptly. This feature allows short-term (months) price predictions.

## 2. **Predicted vs. actual**

Originally, we presented a model for the price evolution for motor fuel and crude oil from March to December 2009. For the motor fuel index, we drew a straight line shown in the left panel of Figure 2. This line said that motor fuel price would be growing faster than the price of all goods and services. Specifically, we predicted that:

> In March 2009, the difference was at the level of +45, i.e. much higher than the level predicted by the new trend. As happened in the past with numerous individual price indices [9,10], such a strong deviation (one might call it "dynamic overshoot") should be compensated in the near future. Without loss of generality, we have restricted the recovery to the trend by the end of 2009. As a result the index for motor fuel should growth by 90 units during the next 9 months, or by 10 units per month. Red filled circles represent the evolution of the difference from April to December 2009. In 2010, the difference may undergo an overshoot in the opposite direction with additional rise in the index for motor fuel.
>
> Translating indices into prices, the rise in the difference by 90 units (from 173 in March to 263 in December) means an increase in price by 50%. Therefore, it is very likely that the price for motor fuel in the beginning of 2010 will be 60% to 70% larger than in March 2009 due to the overshoot.

We have been also tracking the evolution of crude oil price, which obviously affects the price index of motor fuel. As Figure 1 demonstrates, these prices have no one-to-one correspondence and it might be instructive to model them separately. So, for oil price we used an alternative approach and assumed that the price trajectory would continue the pendulum-like motion observed since 2008. This implied the difference between the PPI and the index of crude petroleum should sink below the new trend and stop at the level of -120, which roughly corresponds to $120 per barrel.

So, there were two approaches: 1) motor fuel (oil) price would to be stopped at the new trend and then be evolving along it; 2) oil (motor fuel) price should follow up the observed free pendulum oscillation and penetrate deep below the trend line in a dynamic overshoot. The second approach implies that the price will rebound above the trend. Both predictions are shown in Figure 2 by solid diamonds for the period between March and December 2009.

Figure 2 also presents actual prices for the predicted period. Overall, the evolution of the difference between the CPI and the index of motor fuel follows the predetermined path: from its peak in February 2009 to the new trend line, which was reached in December 2009. In January 2010, the actual time series likely started to align along the new trend. One may expect the next move will be below the trend with small-amplitude short-term (few months) oscillations around the trend in 2010 and 2011. In the long run motor fuel will be losing its pricing power relative to the CPI.

Obviously, the assumption behind the oil price prediction was wrong. There was no free pendulum motion observed beyond June 2009. When reached the new (dashed) trend line, the difference started the alignment along the line. Instead of $120 per barrel oil price has been hovering around $75, as was predicted by the motor fuel model. There is no sign that petroleum price will significantly deviate from the trend in the near future. In March and April 2010, we expect the price to rise above the new trend. It must return below the line by the third quarter, however.

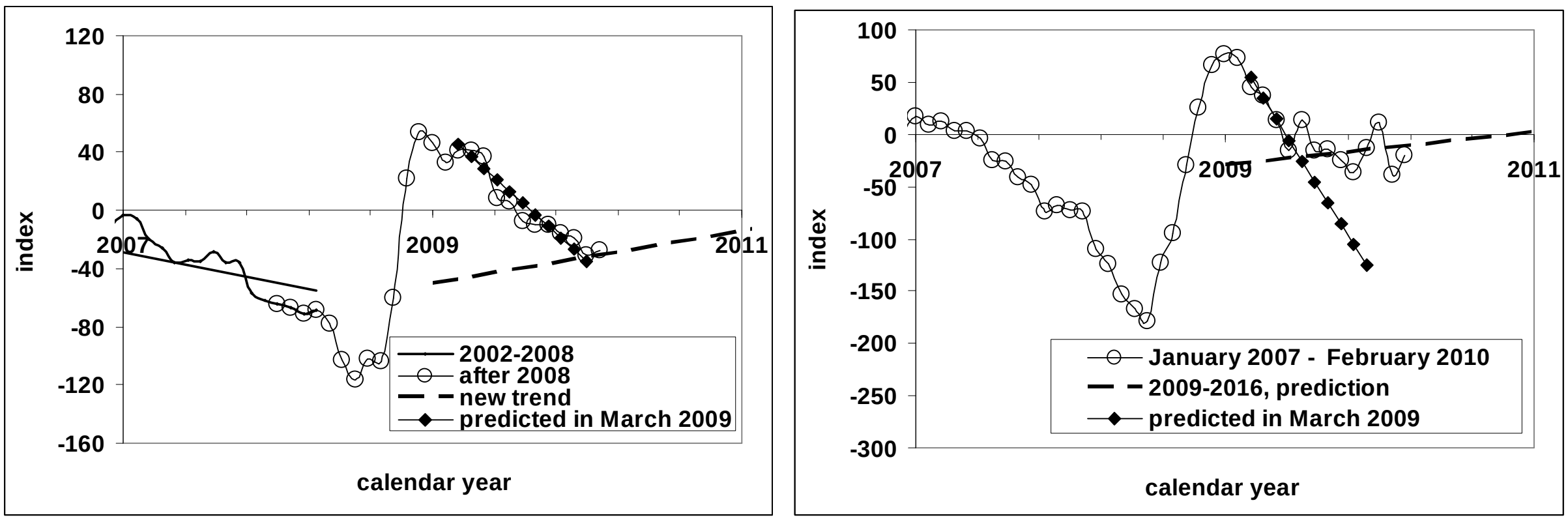

Figure 2. *Left panel*: The difference between the headline CPI and the index for motor fuel. Solid diamonds represent the prediction given in March 2009 through December 2009. The total increase in the difference is +60 units of index or +35%: from 173 in March to 233 in December. Dashed line represents the new trend, which is a mirror reflection to that between 2001 and 2008 shown by solid black line. *Right panel*: Evolution of the difference between the PPI and the index for crude petroleum (domestic production). Solid diamonds represent the predicted values between March and December 2009, as anticipated in April 2009 with a dynamic overshoot below the new trend. Open circles – the observed difference between September and December 2009. Dashed line represents the new trend as described in the text.

From Figure 2, it is clear that the oil price differnce leads that of motor fuel by six to eight months. It is likely that the index of motor fuel will follow up the trajectory drawn by oil price and fluctuate around its trend with slightly larger amplitude. The predictive power of this assumption will be tested by the end of 2010. We are going to track and report on actual behavior. Our model needs further validation in both short- and long run.

## 3. Discussion

All in all, our prediction of the price index of motor fuel was based on a sound assumption and thus is accurate. The forces behind the observed long- and short-term behavior are not accessible yet but very powerful. We dare say they are fundamental and affect the economy to its deepest roots. These forces retain equilibrium among all economic agents and originate the sustainable trends in the differences between consumer (producer) price indices. At some point, the forces meet their limits and should be re-balanced in order not to harm the economy. As a result, the sustainable trends in the CPI and PPI turn.

Meanwhile, it is instructive to revise our long-term prediction of oil price shown in Figure 1. After a few minor adjustments to the initial and final levels of the PPI and the index of crude petroleum, Figure 3 depicts the revised prediction after 2010. It is very similar to the previous prediction with oil price in 2016 set at $30 per barrel. The index of motor fuel will follow up the trend with a delay and larger amplitude. Short-term fluctuations can not be predicted at a horizon of several years. However, the larger is a given deviation from the trend the larger is the returning force.

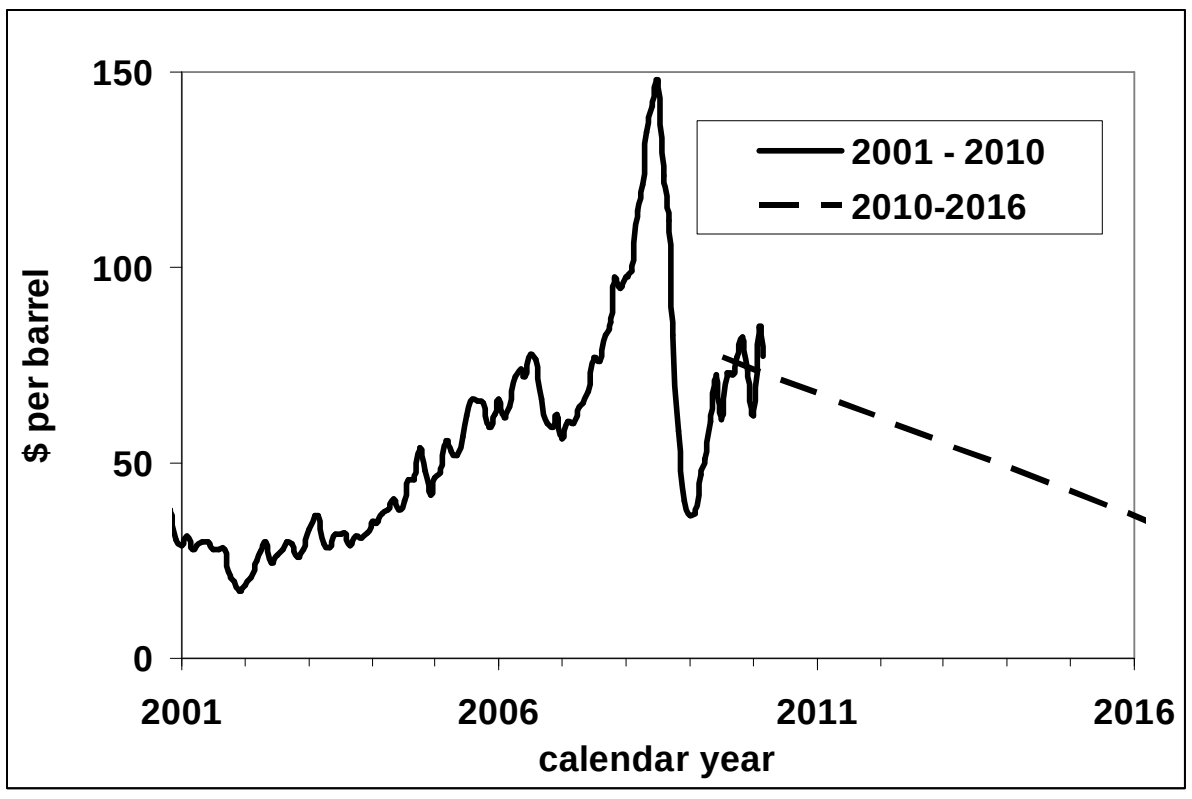

Figure 3. The evolution of crude oil price. Solid line – oil price for the period between 2001 and 2010. Dashed line – the new developed trend between 2009 and 2016. According to the prediction, the price should fall to the level of $30 per barrel by 2016.